\documentclass[
preprintnumbers,
prd,
a4paper,
twocolumn,
amsmath, amssymb, nofootinbib
]{revtex4}
\usepackage[dvips]{graphicx}

\begin{document}

\title{Gravitational collapse disturbs the dS/CFT correspondence?}

\author{Yukinori Iwashita$^{(1)}$, Hirotaka Yoshino$^{(1)}$ and Tetsuya Shiromizu $^{(1,2,3)}$}


\affiliation{$^{(1)}$Department of Physics, Tokyo Institute of Technology, Tokyo 152-8551, Japan}

\affiliation{$^{(2)}$Department of Physics, The University of Tokyo,  Tokyo 113-0033, Japan}

\affiliation{$^{(3)}$Advanced Research Institute for Science and Engineering, 
Waseda University, Tokyo 169-8555, Japan}

\date{\today}

\begin{abstract}
We study the gravitational collapse 
in five-dimensional de Sitter (dS) spacetime and discuss 
the existence of the conformal boundaries at future timelike infinity($\mathcal{I}^+$) 
from the perspective of the dS/CFT correspondence. We investigate 
the motion of a spherical dust shell and the black hole area bounds. 
The latter includes the analysis of the trapping horizon 
and the initial data with spindle-shaped matter distribution. 
In all above analyses we find the evidences that guarantee the existence of 
the conformal boundaries at future timelike infinity 
which may be essential to apply the dS/CFT correspondence. 
\end{abstract}


\maketitle

\label{sec:intro}
\section{Introduction}

A lot of studies of the de Sitter (dS) spacetimes 
have been done so far, mainly motivated by the inflation in the very early universe. 
Recent observations also indicate that our universe has a small and 
positive cosmological constant $\Lambda$ \cite{obss,obsw}. 
This means that our universe would asymptote to the dS. 
The dS has the following interesting properties: it possesses the cosmological horizons 
and accordingly the conformal boundaries exist in its timelike infinity. 
As well as the black hole horizon, the thermodynamics of 
the cosmological horizon has been studied \cite{th1,th2}. 
In principle the description of thermodynamics needs the knowledge of
the microscopic physics. However, the existence of the cosmological horizons 
means the existence of the unobserved spacetime region and 
this fact makes it difficult to understand the origin of the entropy of dS spacetimes.

One of the successful approaches to overcome this difficulty is  
the holography. The most striking example of the holography is
the anti-de Sitter/conformal field theory (AdS/CFT) correspondence 
conjectured by Maldacena \cite{mal}. It states that type IIB string theory on 
$AdS_5\times S^5$ and $\mathcal{N}=4$, $SU(N)$ four dimensional 
Yang-Mills theory are equivalent. The equivalence includes the operator 
observable, the correlation function and the full dynamics. 
Inspired by the AdS/CFT, the dS/CFT correspondence was proposed 
by Strominger \cite{and}. 
The dS/CFT correspondence states that the dual description of the gravity
in the dS spacetimes is the CFT in the future and past conformal boundaries. 
This is intuitively explained as follows: 
the dS spacetime is obtained by the double Wick rotation of the AdS spacetime 
and hence the similar arguments are valid for the asymptotic structure 
and the holography. Although it has no grounds in the string theory, 
there are many strong evidences \cite{dc1,dc2} which indicate the validity of
the dS/CFT correspondence. 
The impressive example is that the bulk scalar field corresponds to 
the CFT on the conformal boundary. 
Then this conjecture motivates us to study the properties of the higher dimensional dS spacetimes.  
Furthermore, the dS/CFT helps us to understand the deSitter entropy\cite{en1,en2}.

One of the interesting issues in these contexts is the dual description of 
the gravitational phenomena caused by the matter in the (A)dS spacetimes 
(see, e.g., \cite{GN02} for the successful boundary description of the matter 
in the AdS case). 
However, in the dS/CFT case it is pointed out 
that the presence of matter drastically alters the causal structure \cite{bk}. 
If a large amount of matter is present, the collapse of matter might cause 
the gravitational collapse. If there is no upper bound on the area of the black holes 
produced in this collapse, it can have the area larger than that of the cosmological 
horizon. This means the collapse of the whole spacetime. If this is the case, 
future timelike infinity is altered by a spacelike singularity 
and there is no conformal boundary on which the dual 
CFT exists (see also \cite{SI} for similar issues in AdS/CFT correspondence 
and braneworld context). This implies the serious problem for the dS/CFT correspondence. 
Hence the role of the matter in the dS spacetimes 
should be investigated in detail. 

Note that the dS has two conformal boundaries($\mathcal{I}^{\pm}$) and 
it may be controversial whether the dS/CFT correspondence involves a single dual CFT or two. 
In this paper we take the latter position as Refs. \cite{en1,dc3}. 
Because we are now interested in the dS dynamics, 
it would be better to consider both past and future timelike infinities together. 

The related subject in four-dimensional dS spacetimes 
is the cosmic no-hair conjecture motivated by the inflationary universe. 
It states that all expanding universes with positive $\Lambda$ 
asymptotically approach the dS spacetimes for local observers \cite{th1,nh}. 
This indicates the existence of the conformal boundary at timelike infinity: 
even if the matter collapses, it does not cause the collapse of whole spacetime.  
In four dimensions cosmic no-hair conjecture was tested in the several 
situations and its validity has been confirmed 
except for some particular examples~\cite{enh1,enh2}. 
Furthermore, the area bound on the apparent horizon was discussed 
in \cite{area1,area2}, supporting the cosmic no-hair conjecture. 
Thereafter the shell motion was also studied in \cite{nakao1,nakao2}.

In this paper we will examine some aspects of the gravitational 
collapse in five-dimensional dS spacetimes 
in order to obtain implications about the generality of 
the existence of conformal boundary at future timelike infinity 
in higher dimensions, 
which is essential to apply the dS/CFT correspondence. 
We shall address this problem by two ways. 
One is to analyse the motion of shell-shaped matter 
in dS spacetime and its gravitational collapse. 
This would provide us some informations about 
the matter collapse in dS spacetime. 
The other is to discuss the area bounds on the
black holes in the dS spacetimes because such bounds strongly indicates that the matter 
does not cause the collapse of the whole spacetime. 
We investigate the spherically-symmetric vacuum case and study 
the area of trapping horizons in rather general way. 
We will also study 
the area bounds on the apparent horizon numerically, setting 
the initial data generated by the spindle-shaped matter source 
in the dS spacetime. 
We would like to remark that the arguments in four dimensions 
can not be straightforwardly extended to higher dimensions, 
since the argument is often relied on the four dimensional speciality.

This paper is organised as follows. In the next section 
we study the motion of spherical dust shell. 
In Sec. III 
the area bounds on the event horizons and the apparent horizons are
investigated using various examples. 
The Schwarzschild-deSitter black holes, 
trapping horizons in more general setting and the apparent horizons 
in the initial data are studied. 
Finally we summarise our results in section IV. 
We use the unit $c=G=1$ throughout this paper.


\section{Shell motion in dS spacetimes}

In this section, we study the dynamical motion of a spherically symmetric dust shell. 
This is the five-dimensional generalisation of the study in \cite{nakao1}. 
By this analysis we might have some insights into the existence of the conformal boundary
at future timelike infinity.

Let $\Sigma$ denote the trajectory of the thin shell which is the
timelike hypersurface in the spacetime and divides the spacetime into   
the inside region $V^-$ and the outside region $V^+$. 
Let $n^a$ be a unit normal to $\Sigma$ and we introduce the intrinsic metric 
$h_{ab}=g_{ab}- n_a n_b$ 
and the extrinsic curvature $K_{ab}=h_a^c \nabla_{c} n_{b}$. 
We define
%
\begin{eqnarray}
	[A]&=& A^+-A^-, \\
	\{A\}&=& A^++A^-:=\bar{A}.
\end{eqnarray}
%
$+$ and $-$ denote the tensor in the region $V^+$ and $V^-$ respectively. The junction condition is 
%
\begin{eqnarray}
	[K_{ab}]&=& -8\pi \left( S_{ab}-\frac{1}{3}h_{ab}S\right), \label{jc}
\end{eqnarray}
%
$S_{\mu\nu}$ is obtained by
%
\begin{eqnarray}
	S_{\mu\nu}=\lim_{\epsilon\rightarrow 0} \int^{+\epsilon}_{-\epsilon}T_{\rho\sigma}h^{~\rho}_{\mu}h^{~\sigma}_{\nu}dx,
\end{eqnarray}
%
where $x$ is a Gaussian coordinate orthogonal to $\Sigma$.
Using the Gauss-Codacci equations we obtain the following relations,
%
\begin{multline}
	{}^{(4)}R + \bar{K}_{ab} \bar{K}^{ab}-\bar{K}^2 = -16\pi \left( S_{ab}S^{ab}-\frac{1}{3}S^2\right)\\ 
	-8\pi \{ T^{ab}n_a n_b\}, \label{hc} 
\end{multline} 
\begin{align}
	\bar{K}_{ab}S^{ab}&=[T^{ab}n_a n_b], \label{d1}\\
	D_b\bar{K}_a^{~b}-D_a\bar{K} &= 4\pi \{ T^{bc}n_b h_{ca}\}, \label{mc}\\
	D_bS_a^{~b} &= -[T^{bc}n_b h_{ca}].\label{d2}
\end{align}
%
Equations \eqref{hc} and \eqref{mc} is the Hamiltonian constraint 
and the momentum constraint respectively 
while Eqs. \eqref{d1} and \eqref{d2} are the evolution equations of the shell motion. 

Assuming that the inside of $\Sigma$ is empty, 
$V^-$ should be the dS spacetime and 
$V^+$ the Schwarzschild-dS spacetime. This metric is written as
%
\begin{eqnarray}
	ds^2=-f_{\pm}(r_{\pm})dt_{\pm}^2+f_{\pm}^{-1}(r_{\pm})dr_{\pm}^2+r_{\pm}^2 d \Omega^2_3,
\end{eqnarray}
%
where $f_{\pm}$ are given by
%
\begin{eqnarray}
	f_{-}&=&1-r^2/\ell^{2}\\
	f_{+}&=&1-r^2/\ell^{2}-2M_g/r^{2},
\end{eqnarray}
%
where $\ell$ is the dS radius $\ell\equiv\sqrt{6/\Lambda}$ and $M_g$
is related to the black-hole gravitational mass $m_g$ by $M_g=4m_g/3\pi$ .
The metric on $\Sigma$ is written as
%
\begin{eqnarray}
	dl^2= -d\tau^2+R^2(\tau)d\Omega^2_3
\end{eqnarray}
%
where $\tau$ is the proper time of the dust shell and $R(\tau)=r_{\pm}(\tau)|_{\Sigma}$ 
is the radius of the dust shell. 
The non-zero components of the extrinsic curvature are 
%
\begin{eqnarray}
	K^{\tau}_{\pm\tau}&=& -\sigma_{\pm}\left( \ddot{R}+\frac{1}{2}\frac{df_{\pm}}{dR} \right)\Big/ \sqrt{\dot{R}^2+f_{\pm}} \\
	K^{\theta_1}_{\pm\theta_1}&=& K^{\theta_2}_{\pm\theta_2}=K^{\theta_3}_{\pm\theta_3}
	= -\sigma_{\pm} \sqrt{\dot{R}^2+f_{\pm}}\Big/ R,
\end{eqnarray}
%
where $\sigma_{\pm}=\pm 1$. For the dust shell, 
%
\begin{eqnarray}
	S_{\tau}^{~\tau}=-\rho
\end{eqnarray}
%
holds. From Eq. (\ref{d2}) we obtain the conservation law
%
\begin{eqnarray}
	\frac{d(\rho R^3)}{d\tau}=0.
\end{eqnarray}
%
This means that the proper mass of the dust shell $m_s\equiv 2\pi^2\rho R^3$ is 
the conserved quantity. For convenience, we introduce $M_s\equiv 4m_s/3\pi$.

Eqs. \eqref{d1} and \eqref{jc} yield the equation of motion 
%
\begin{eqnarray}
	\ddot{R}=-\frac{M_g}{R^3}-\frac{(8\pi )^2}{18}\rho^2 R+\frac{R}{\ell^2}.
\end{eqnarray}
%
Integrating with $\tau$, we obtain
%
\begin{equation}
	\frac{1}{2}\dot{R}^2=-V(R)\equiv \frac{M_g}{2R^2}+\frac{M_s^2}{4 R^4}+\frac{R^2}{2\ell^2}  
	+ \frac{M_g^2}{M_s^2} -1. \label{pot}
\end{equation}
%
The region $V(R)>0$ is forbidden in the sense
that the recollapse / the bounce occur at $V(R)=0$ and the shell cannot take 
such $R$. $V(R)$ has the maximum 
%
\begin{eqnarray}
	V(R_0) = -\frac{M_g}{4R_0^2}-\frac{3R_0^2}{4\ell^2} 
	+\left( 1- \frac{M_g^2}{M_s^2} \right),
	\label{V_maximum}
\end{eqnarray}
%
where
%
\begin{eqnarray}
	R_0^2 = X_{+}^{{1}/{3}} + X_{-}^{{1}/{3}}, 
\end{eqnarray}
%
%
\begin{eqnarray}
	X_{\pm} = \frac12 M_s^2\ell^2
	\left( 1\pm \sqrt{1-\frac{4\ell^2 M_g^3}{27M_s^4}} \right).
\end{eqnarray}
%

\begin{figure}[tb]
	\begin{center}
		\includegraphics[width=5.5cm,angle=-90]{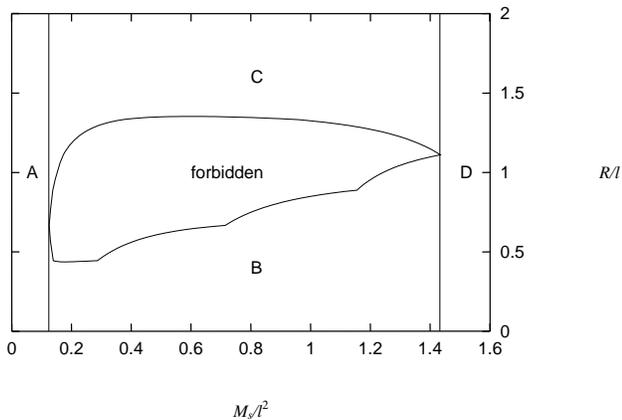}
	\end{center}
	\caption{
	The allowed and forbidden region for the case $M_g/\ell ^2=0.1$. 
        The allowed region is composed of four parts, $A, B, C$ and $D$. } 
	\label{shell}
\end{figure}

Now we look at the details of the forbidden and allowed regions in the $(M_s/\ell ^2, R/\ell)$-plane fixing $M_g/\ell ^2$.
The case $M_g/\ell ^2=0.1$ is shown in Fig.~\ref{shell}. 
The allowed region is composed of four parts, $A, B, C$ and $D$. 
As we can see easily by Eq.~\eqref{V_maximum},  there is no forbidden region 
for $M_g\gtrsim M_s$. 
This is depicted as the region $A$ Fig.~\ref{shell}. It would be surprising that 
the forbidden region also does not exist for sufficiently large $M_s$ 
as depicted by the region $D$. 
In these regions $A$ and $D$ the shell continue to collapse or
expand depending on the initial condition. The spacetime
approaches the dS spacetime if the shell expands and 
does the Schwarzschild-dS spacetime if the shell collapses. 
In the region $B$, due to the existence of
the forbidden region, the shell eventually collapses,
even if it initially expands. The spacetime approaches the Schwarzschild-dS
spacetime. In the region $C$ the converse statement holds:   
the shell bounces even if it initially collapses and the spacetime evolves into 
the dS spacetime.

The important lesson from this result is as follows. In the above analysis 
one can choose arbitrary values of the proper mass $M_s$ for
a fixed gravitational mass $M_g$. One would expect that
$M_g-M_s$ provides the binding energy and the system
would be bounded if it is negative. However, 
$M_s$ has the maximum value where recollapse occurs and 
the shell is not bounded for larger $M_s$ values, which
is contradictory to our intuition.  
This is because large $M_s$ enlarges the second term of Eq. \eqref{pot} 
and suppress the maximum of the potential. 
Hence, we have an expectation
that even there is large mass distribution in the dS spacetime,
it would not be gravitationally bounded and would not cause the collapse of the
whole spacetime.

In the next section we address this problem
from somewhat different point of view:
the bounds on the black-hole area in the dS spacetime.

\section{Area bounds}
\label{sec:example}

In this section we study the area bound on the black holes in five dimensional 
dS spacetimes. As we mentioned in Sec. I, if 
the black-hole area has the upper bounds, 
it strongly indicates that the collapse of the whole spacetime 
does not occur even if the heavy matter exists.

\subsection{Schwarzschild-dS spacetime}

We begin with the Schwarzschild-dS solution. 
The area bounds derived in this analysis is expected to have 
the generality because the Schwarzschild-dS is the most symmetric configuration 
as  mentioned in \cite{area1}.
The metric is 
%
\begin{eqnarray}
ds^2=-f(r)dt^2+f^{-1}(r)dr^2+r^2 d \Omega^2_3,
\end{eqnarray} 
%
%
\begin{eqnarray}
f(r)=1-\frac{2M}{r^2}-\frac{r^2}{\ell^2}, 
\end{eqnarray}
%
where $M$ is related to the gravitational mass $m$ by $M=4m/3\pi$. 
$f(r)=0$ has two positive roots. The smaller one represents the location
of the black hole event horizon
%
\begin{eqnarray}
r^{\rm (BH)}=\ell\left[\left({1-\sqrt{1-8M/\ell^{2}}}\right)\big/{2}\right]^{1/2},
\end{eqnarray}
%
and the other gives that of the cosmological event horizon
%
\begin{eqnarray}
r^{\rm (CH)}=\ell\left[\left({1+\sqrt{1-8M/\ell^{2}}}\right)\big/{2}\right]^{1/2}.
\end{eqnarray}
%
If $M=\ell^2/8$, $r^{\rm (BH)}$ and $r^{\rm (CH)}$ coincide.
If we further increase $M$,  
there are no positive roots for $r^{\rm (BH)}$ and $r^{\rm (CH)}$.
Therefore $r^{\rm (BH)}$ has the maximum value and 
accordingly the black hole mass and area have the following upper bounds:
%
\begin{eqnarray}
M \leq M_c:=\frac{1}{8}\ell^2 ,
\end{eqnarray}
%
%
\begin{eqnarray}
A^{\rm (BH)} \leq A_c:= \frac{\pi^2}{\sqrt{2}} \ell^3.\label{Area_bound_spherical}
\end{eqnarray}
%
This suggests that the area of any black holes in asymptotically dS spacetimes 
is bounded above by $A_c$. 
Hence the formation of a large black hole is expected to be prohibited. 
This supports the existence of the conformal boundary at future infinity. 
To study this expectation further, we consider the trapping horizon in the
next subsection,
of which area bounds we can discuss more generally.

\subsection{The trapping horizon}

In this subsection 
we discuss the trapping horizons 
in five-dimensional dS spacetimes following the procedure in \cite{area1} 
without assuming the spherical symmetry. 
The existence of the trapping horizon is the 
sufficient condition for the event horizon formation, 
and the area bounds of the trapping horizon have implications 
about the black-hole area bounds.

Let $\xi_{\pm}$ be the double-null coordinates such that
the three-dimensional surface $S$ given by $\xi_{\pm}={\rm const.}$ is closed.  
We introduce the normal 1-forms $n_{\pm a}=-d\xi_{\pm a}$, their dual $N_{\pm}^a=g^{ab}n_{\pm b}$ 
and $u_{\pm}^a=(\partial /\partial \xi_{\pm})^a$.
The induced metric on $S$ is written as
%
\begin{eqnarray}
	q_{ab}=g_{ab}+2e^{-f}n_{+a}n_{-b},
\end{eqnarray}
%
where $e^f = -g_{ab}N_+^aN_-^b$. 
The shift vectors are $r_{\pm}^a=q^{ab}{}^{(3)}h_{bc}u_{\pm}^c$ 
and we denote the Lie derivatives along $u_{\pm}^a-r_{\pm}^a$ by $\mathcal{L}_{\pm}$.
The expansions $\theta_{\pm}$, shears $\sigma_{\pm ab}$, inaffinities $\nu_{\pm}$ 
and twists $\omega_a$ are defined by
%
\begin{eqnarray}
	\theta_{\pm}&=& \frac{1}{2}q^{ab}\mathcal{L}_{\pm}q_{ab},\label{exp}\\
	\sigma_{\pm ab}&=& q_a^{~c}q_b^{~d}\mathcal{L}_{\pm}q_{cd}
	-\frac{1}{3}q_{ab}q^{cd}\mathcal{L}_{\pm}q_{cd},\\
	\nu_{\pm} &=& \mathcal{L}_{\pm}f,\\
	\omega_a &=& \frac{1}{2}e^{-f}q_{ab}[N_+, N_-]^b.
\end{eqnarray}
%
Using the Einstein equation, we obtain the following cross-focusing equations
%
\begin{multline}
\mathcal{L}_- \theta_+ + \theta_+ \theta_- +e^{-f}\left[(1/2){}^{(3)}R - \tau_a \tau^a -
{\cal D}_a \tau^a  \right]  \\
= 8\pi \rho + 6e^{-f}\ell^{-2}, \label{cro}
\end{multline}
%
where $\rho$ is given by $\rho =T_{ab}(u_+^a-r_+^a)(u_-^b-r_-^b)$
with the energy-momentum tensor $T_{ab}$, 
$\tau_a=\omega_a-({1}/{2})\mathcal{D}_a f$,
${}^{(3)}R$ denotes the Ricci scalar on $S$  
and $\mathcal{D}_a$ denotes the covariant derivative associated with $q_{ab}$.

The marginal surface is defined as a spatial three-dimensional surface $S$ 
on which null expansion vanishes, $\theta_+|_S= 0$. 
The future, outer, trapping horizon is the four-dimensional surface $H$ such 
that it is foliated by the marginal surfaces $S$  on which further conditions
$\theta_-|_S< 0$ and $\mathcal{L}_-\theta_+|_S < 0$ are satisfied.
Now let us integrate Eq. \eqref{cro} over $S$ of the trapping horizon. 
If we assume the dominant energy condition $\rho \geq 0$ and $\int_S {}^{(3)}R dA > 0$, 
we obtain the inequality 
%
\begin{equation}
A_3 \leq  \frac{\ell^2}{12}\int_S {}^{(3)}R dA. \label{hs}
\end{equation}
%
where $A_3$ is the three-dimensional area of $S$. 
Hence the area $A_3$ is bounded by the integral of the Ricci scalar
of the surface $S$. This would be regarded as the manifestation
of the black-hole area bounds in five-dimensional dS spacetimes.

In four dimensional cases, it is shown that the foliation $S$ of the 
trapping horizon has spherical topology \cite{SH} 
using the Gauss-Bonnet theorem. As a result,  
the area of the spatial section of the trapping horizon 
is bounded by that of the cosmological horizon \cite{area1}. 
However, in the higher-dimensional cases we cannot use the 
Gauss-Bonnet theorem or conclude 
such bounds in general. 
Therefore we numerically
study the area bounds further with a concrete example
in the next subsection.

\subsection{The spindle initial data }

In this subsection we study the area bounds on the apparent horizon
by analyzing the four-dimensional initial data in a five-dimensional
spacetime with positive $\Lambda$.
Because the existence of the apparent horizon is the sufficient condition
of the black hole formation,  the area bounds on the apparent horizon
gives some implications about those on the black hole.
As the most simple example of the system without spherical symmetry 
we consider the uniform line-shaped distribution of matter.
This is the extension to the space with positive $\Lambda$
of the study in \cite{in}.
We also observe the area of the ``cosmological horizon'', which will
be introduced later.

Let us consider spacelike hypersurface $\Sigma$
with the metric $h_{ab}$ and the extrinsic curvature $K_{ab}=-h_a^c\nabla_cn_b$.
The Hamiltonian constraint is 
%
\begin{eqnarray}
	{}^{(4)}R+K^2-K_{ab}K^{ab}= 12 \ell^{-2},\label{hads}
\end{eqnarray}
%
where ${}^{(4)}R$ is the Ricci scalar on $\Sigma$. 
We adopt the extrinsic curvature, which is the solution of the
momentum constraint, as follows:
%
\begin{eqnarray}
	K_{ab} =  -\frac{1}{\ell} h_{ab}.
\end{eqnarray}
%
This extrinsic curvature indicates that all points in $\Sigma$ 
is expanding with the same rate. Substituting this 
formula into Eq. \eqref{hads}, we obtain
%
\begin{eqnarray}
	{}^{(4)}R=0. \label{initial}
\end{eqnarray}
%
Assuming the conformal flatness $h_{ab}=\varPsi^2\delta_{ab}$,
Eq.~\eqref{initial} becomes the Laplace equation:
%
\begin{eqnarray}
	\nabla^2\varPsi=0.
\end{eqnarray}
%
Now we introduce the cylindrical coordinate $(z, \rho)$, in which the
metric becomes
%
\begin{eqnarray}
	ds^2 = \varPsi^2(\rho ,z)[dz^2+d\rho^2+\rho^2d\Omega_2^2],
\end{eqnarray}
%
where $d\Omega_2^2$ means the metric on the unit two-dimensional sphere,
and adopt the solution given in \cite{in}:
%
\begin{eqnarray}
	\varPsi = 1+\frac{2m}{3\pi L\rho} \left( \arctan \frac{z+L/2}{\rho}-\arctan\frac{z-L/2}{\rho} \right).
\end{eqnarray}
%
This solution represents the space with the a uniform line source of the length $L$ 
distributed at $z$-axis. 
We obtain the flat case by taking the limit $\ell \to \infty$ and the spherical case by $ L \to 0$.

To determine the location of the apparent horizon that surround the line source, 
we introduce a new coordinate ($r, \theta$) by 
%
\begin{eqnarray}
	\rho = r \sin \theta ~~{\rm and}~~z= r \cos \theta. 
\end{eqnarray}
The apparent horizon is the surface on which the
expansion $\theta_+$ defined by Eq. \eqref{exp} vanishes.
We assume that the apparent horizon is located 
at $r=h(\theta)$ and introduce the unit normal $s^a$ to this surface.
$\theta_+ =0$ is rewritten as
%
\begin{eqnarray}
	D_as^a-K+K_{ab}s^as^b=0.\label{AH_eq}
\end{eqnarray}
%
$s^a$ is given by
%
\begin{equation}
s^a=\frac{1}{\varPsi\sqrt{1+h_{,\theta}^2/r^2}}(1,-h_{,\theta}/r^2),
\end{equation}
%
and we obtain the equation for $h(\theta)$ by substituting to Eq. \eqref{AH_eq}
as follows:
%
\begin{multline}
	h_{,\theta\theta}-3\left( \frac{\varPsi_{,r}}{\varPsi}+\frac{1}{h} \right)h^2 - 
	\left( 3\frac{\varPsi_{,r}}{\varPsi}+\frac{4}{h} \right)h_{,\theta}^2 \\
	+\left( 3\frac{\varPsi_{,\theta}}{\varPsi}+2\cot\theta \right)
	h_{,\theta} \left( 1+\frac{h_{,\theta}^2}{h^2} \right)\\
	-\frac{3}{\ell} \frac{\varPsi}{h} \left( h^2+h^2_{,\theta} \right)^{3/2} =0. \label{mots}
\end{multline}
%
We solve this equation numerically
under the boundary condition $h_{,\theta}(0)=h_{,\theta}(\pi/2)=0$.
The area $A_3^{\rm (AH)}$ of the apparent horizon is calculated by
\begin{equation}
A_3^{\rm (AH)}=8\pi\int_0^{\pi/2}\varPsi^{3}h^2\sqrt{h^2+h_{,\theta}^2}d\theta.
\label{area_formula}
\end{equation}

If we change $\ell$ to $-\ell$ in Eq.~\eqref{mots}, 
the rewrote equation gives the solutions for $\theta_-=0$. 
Outermost solution of $\theta_-=0$ 
corresponds to the cosmological horizon in the spherically symmetric
case($L=0$). 
Hence we simply call it a ``cosmological horizon'' hereafter.
The area $A_3^{\rm (CH)}$ of the ``cosmological horizon''
is calculated by the same formula as Eq. \eqref{area_formula}.

For numerical convenience 
we used $\sqrt{2m/3\pi}$ as the unit of the length in this analysis.
In other words, we fixed $m$ and changed the values of $L$ and $\ell$. 
We will investigate the existence and the area of both the apparent horizon
and ``cosmological horizon'' changing the values of $L$ and $\ell$.

\begin{figure}[tb]
	\begin{center}
		\includegraphics[width=5.5cm,angle=-90]{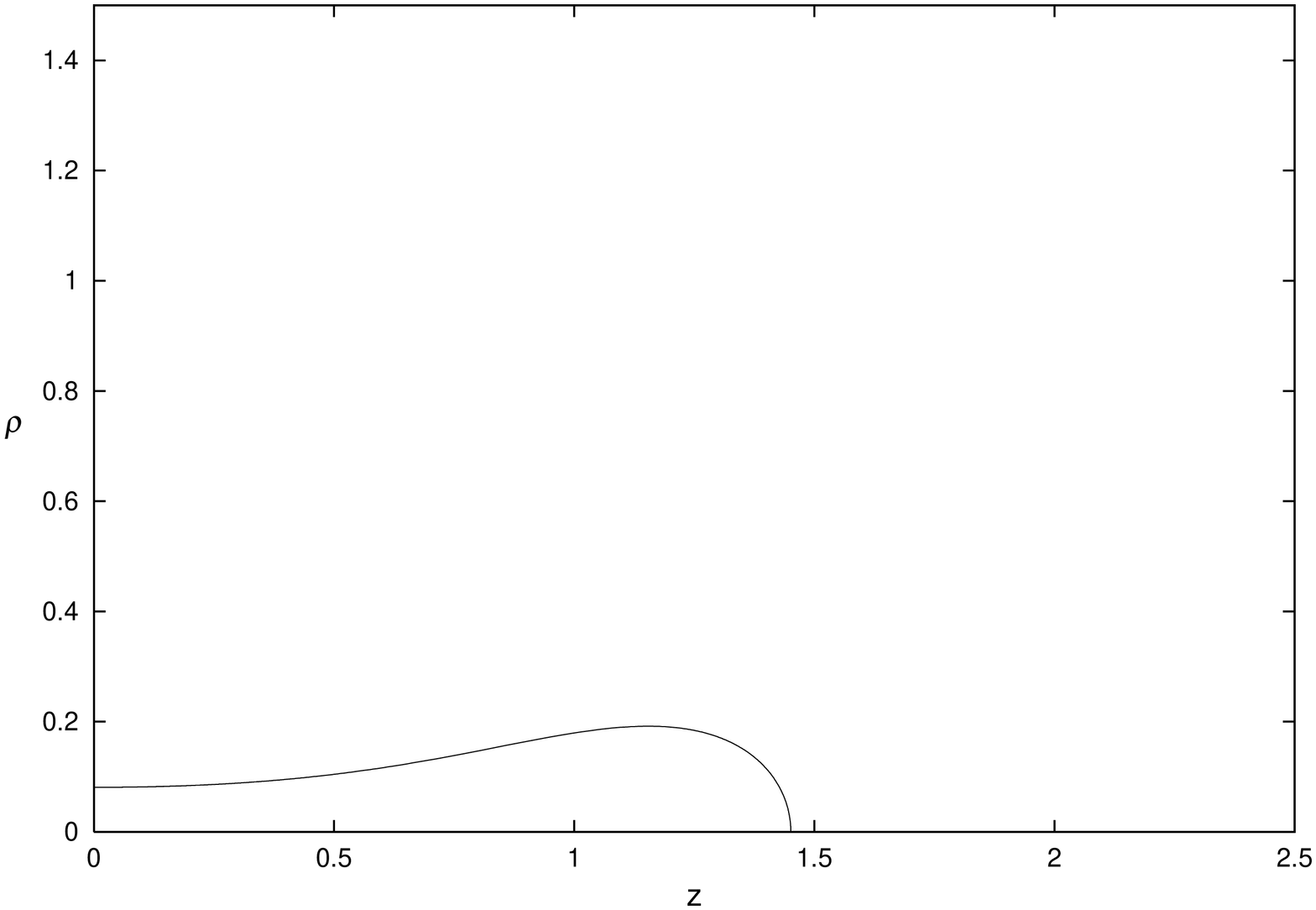}
	\end{center}
	\caption{The shapes of the apparent horizon for $L$=2.7 $\ell$=3.9. 
	The apparent horizon exists, while the ``cosmological horizon'' does not exist.} 
	\label{39}
		\begin{center}
		\includegraphics[width=5.5cm,angle=-90]{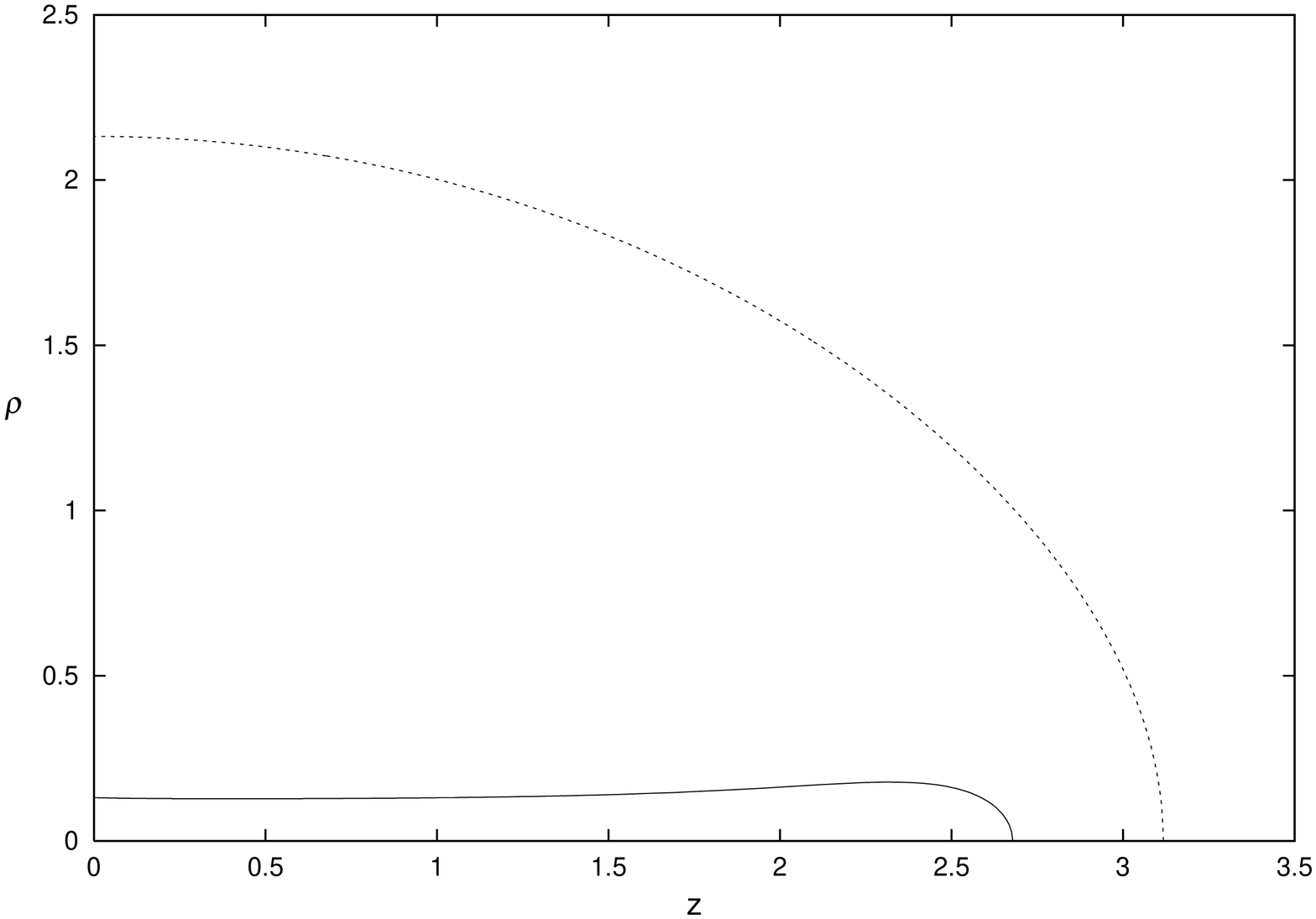}
	\end{center}
	\caption{The shapes of the horizons for $L$=5.2 $\ell$=3.9. 
	Both apparent horizon (solid line) 
	and "cosmological horizon'' (dotted line) exist.} 
	\label{52}
\end{figure}

\begin{figure}[tb]
	\begin{center}
		\includegraphics[width=5.5cm,angle=-90]{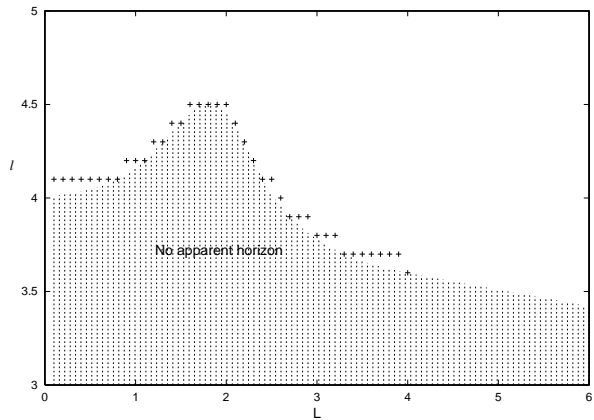}
	\end{center}
	\caption{The region in ($L,\ell$) plane where the apparent horizon exists. 
                    The dotted region represents the region of no apparent horizon. 
                    The crosses represent the data points for which we could find the 
                    apparent horizons.} 
	\label{region}
\end{figure}

The typical shapes of the two horizons are shown in Figs. \ref{39} and \ref{52}
for $\ell=3.9$. 
For this value of $\ell$, both the apparent horizon and the ``cosmological horizon''
do not exist in the spherically symmetric case $L=0$. However, for $L\simeq 2.7$, 
the solution $\theta_+=0$ appears. If we further increase
$L$, the solution of the ``cosmological horizon'' also appears at $L\simeq 5.2$. 
The region in ($L,\ell $) plane where the apparent horizon exists is shown in Fig.~\ref{region}. 
The apparent horizon always exists for large $L$. On the other hand,
for $\ell\lesssim 4.5$, we find the cases
where there is no apparent horizon depending on the value of $L$. 
This result is interpreted as follows. For the large $L$ value,
the typical length scale of the system is $L$ in the parallel direction of
the line source and $m/L$ in the transverse direction of the line source.
The gravitational field in the transverse direction 
becomes almost four dimensional and it does not ``feel'' the curvature of
the space caused by $\Lambda$ 
if $m/L\ll \ell$. Hence, for $L\gg m/\ell$, the long-shaped
apparent horizon can form even if $\ell<4$.

Now we look at the area of two horizons $A_3^{\rm (AH)}$ and $A_3^{\rm (CH)}$. 
Our main interests are whether the area bound in the spherically symmetric
case, Eq.~\eqref{Area_bound_spherical} is satisfied for this highly 
deformed system, and whether the area of the apparent horizon
is bounded by that of the ``cosmological horizon''.

\begin{figure}[tb]
	\begin{center}
		\includegraphics[width=5.5cm,angle=-90]{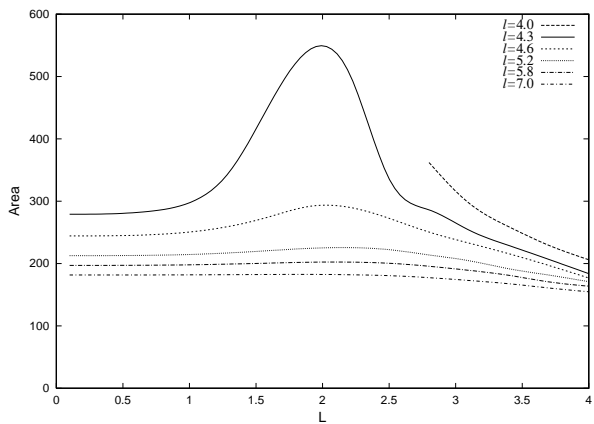}
	\end{center}
	\caption{The area of the apparent horizon as functions of $L$ 
	for $\ell=4.0,~4.3,~4.6,~5.2,~5.8$ and $7.0$. 
	Area has the extremum for $\ell\gtrsim 4.3$ and monotonously decease for large $L$.
	For $\ell\lesssim 4.3$ there is no extremum and the function is defined only
	for large $L$. } 
	\label{arap}
\end{figure}

Fig.~\ref{arap} shows the behavior of $A_3^{\rm (AH)}$ 
as functions of $L$ for several values of $\ell$.
For $\ell\gtrsim 4.3$ there is the extremum of the area and the extremum
value becomes larger for smaller $\ell$. 
For $\ell \lesssim 4.3$ the extremum disappears 
reflecting the existence of the $L$ values for which the apparent horizon
does not exist. We calculated the value of $A_3^{\rm (AH)}/A_c$
and found that ${A_3^{\rm (AH)}}/{A_c} < 1$ is always satisfied. For example, 
in the $\ell=4.3$ case the maximum value of $A_3$ is $\simeq 550$ and hence
\begin{equation}
{A_3^{\rm (AH)}}/{A_c}\simeq 0.99,
\end{equation}
which is consistent with Eq.~\eqref{Area_bound_spherical}. 
This result strongly indicates that the area bound derived in the
spherically symmetric case also holds for highly deformed systems.

\begin{figure}[tb]
	\begin{center}
		\includegraphics[width=5.5cm,angle=-90]{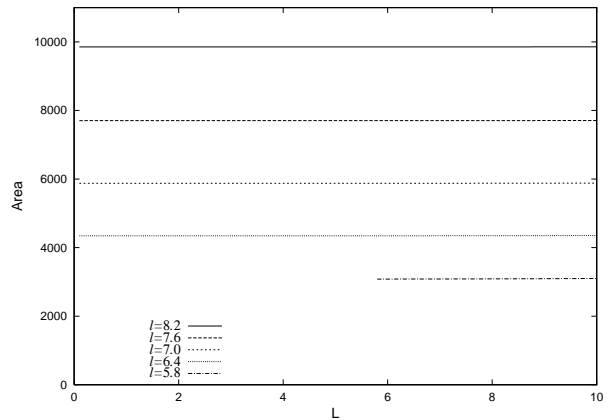}
	\end{center}
	\caption{The area of the "cosmological horizon" as the function of $L$ 
	for $5.8\le\ell\le8.2$ with $0.6$ interval. 
	The area scarcely depend on $L$ if it exists.} 
	\label{arco}
\end{figure}

Now we look at $A_3^{\rm (CH)}$.
Figure~\ref{arco} shows the values of $A_3^{\rm (CH)}$ as
functions of $L$ for several values of $\ell$. 
$A_3^{\rm (CH)}$ has its value for arbitrary $L\ge 0$ for $\ell\gtrsim 6.1$, 
while the $\ell=5.8$ line has the end point at $L \simeq 5.8$,
reflecting the fact that there exists no ``cosmological horizon" for smaller $L$ values. 
The value of $A_3^{\rm (CH)}$
scarcely depends on $L$ and it is almost determined only by $\ell$.
It is obvious that 
\begin{equation}
A_3^{\rm (AH)}<A_3^{\rm (CH)}
\end{equation}
is satisfied. Hence, the area of the apparent horizon is bounded by
that of the ``cosmological horizon''.
Furthermore, we found that $A_3^{\rm (CH)}$ is bounded
below like
\begin{equation}
{A_3^{\rm (CH)}}>A_c,
\end{equation}
which is also consistent with the spherically symmetric case.


\section{Summary}

In this paper we studied the gravitational effect of the matter in 
five-dimensional dS spacetimes.
In Sec. II we analyzed the motion of a spherically symmetric shell
and found that the proper mass of the shell
has the both minimum and maximum values where recollapse occurs
for a fixed gravitational mass. This indicates that the difference
of the gravitational mass and the proper mass does not provide the
binding energy of the system. If the proper mass is much larger
than the gravitational mass, the system is not bounded in contrast to our
intuition. Hence, in the dS spacetime the huge matter distribution
would not cause the collapse of the whole spacetime.

In Sec. III we studied the area bounds on the horizons.
We studied the spherically symmetric case and derived
the area bounds that is expected to hold also in the
general systems. We then
proved the area bounds on the trapping horizon
without spherical symmetry. The bound was 
given in terms of the integral of the Ricci scalar of the horizon.
Although this would be regarded as the manifestation of the area bound,
we could not evaluate it in general. 
Hence we studied the area bound on the apparent horizon
by the initial data analysis. We set the initial data with spindle-shaped matter
in dS spacetime and found that the area of an apparent horizon is bounded 
above by $A_c$ in Eq.~\eqref{Area_bound_spherical}
and also by that of the ``cosmological horizon".
We also found that the area of the ``cosmological horizon'' is
bounded below by $A_c$. All these results indicate that
the area of the black hole is bounded above in five-dimensional dS spacetime. 
Therefore it is unlikely that the matter causes the collapse of the whole spacetime.

Our study supports the existence
of the conformal boundary at future timelike
infinity in the five-dimensional dS spacetime  
even if a large amount of matter exists. 
The results might provide the ground for applying the
dS/CFT correspondence, especially the dual description of 
the gravitational phenomena caused by matter in the dS spacetimes.

%
%

\section*{Acknowledgements}
We acknowledge G. W. Gibbons for helpful comments.
YI and HY thank A. Hosoya and M. Siino for their continuous encouragements. 
The work of TS was supported by Grant-in-Aid for Scientific 
Research from Ministry of Education, Science, Sports and Culture of 
Japan (No.13135208, No.14740155, No.14102004, No. 17740136 and 
No. 17340075).

\end{document}